\documentstyle[preprint,aps,epsfig]{revtex}

\begin{document}

\title{On anharmonicities of giant dipole excitations}

\author{D.T. de Paula\( ^{1} \) , T. Aumann\( ^{2} \), L.F. Canto\( ^{1} \),
B.V. Carlson\( ^{3} \), H. Emling\( ^{2} \) and M.S. Hussein\( ^{4} \) }

\address{\( ^{1} \)Instituto de F\'{\i}sica, Universidade Federal do Rio
de Janeiro, C.P. 68528, 21945-970 Rio de Janeiro, RJ, Brazil\\
 \( ^{2} \) Gesellschaft f\"{u}r Schwerionenforschung (GSI), Planckstr.
1, D-64291 Darmstadt, Germany\\
 \( ^{3} \)Departmento de F\'{\i}sica, Instituto Tecnol\'{o}gico
de Aeron\'{a}utica - CTA, 12228-900, S\~{a}o Jos\'{e} dos Campos,
SP, Brazil\\
 \( ^{4} \)Instituto de F\'{\i}sica, Universidade de S\~{a}o Paulo,
C.P. 66318, 05389-970, S\~{a}o Paulo, SP, Brazil}

\date{\today{}}

\maketitle
\begin{abstract}
The role of anharmonic effects on the excitation of the double giant
dipole resonance is investigated in a simple macroscopic model. Perturbation
theory is used to find energies and wavefunctions of the anharmonic
oscillator. The cross sections for the electromagnetic excitation
of the one- and two-phonon giant dipole resonances in energetic heavy
ion collisions are then evaluated through a semiclassical coupled-channel
calculation. It is argued that the variations of the strength of the
anharmonic potential should be combined with appropriate changes in
the oscillator frequency, in order to keep the giant dipole resonance
energy consistent with the experimental value. When this is taken
into account, the effects of anharmonicities on the double giant dipole
resonance excitation probabilities are small and cannot account for
the well known discrepancy between theory and experiment. 
\end{abstract}
\pacs{ }

The double giant dipole resonance (DGDR) has attracted considerable
interest in the last decade. Several experiments to measure the DGDR
cross section using relativistic heavy ion beams have been performed\cite{emling,Sch93,Aum93,Rit93,Bee93,Bor96}.
Comparison with the predictions of the harmonic oscillator model has
clearly demonstrated a systematic discrepancy. The experimental values
for the DGDR cross sections exceed the theoretical predictions by
a considerable amount. One of the attempts to explain these differences
was made by Bortignon and Dasso\cite{BD97}, using a macroscopic anharmonic
oscillator model. These authors found that with a small anharmonic
perturbation of the \( r^{4} \)-type one can reproduce both the experimentally
observed DGDR excitation energy (which only marginally differs from
that obtained in the harmonic approximation) and the DGDR cross section
for the \( ^{208}Pb+^{208}Pb \) collision at \( 640\,  \)A\( \cdot  \)MeV.
They reached a similar conclusion for the \( ^{136}Xe+^{208}Pb \)
collision at \( 700\,  \)A\( \cdot  \)MeV, where a much greater
discrepancy from the harmonic model appears\cite{Sch93}. The purpose
of this paper is to point out that this model does not lead to the
enhancement found in Ref. \cite{BD97}, if proper renormalization
of the oscillator frequency is performed in order to guarantee that
the theoretical giant dipole resonance (GDR) excitation energy is
kept at the experimental value.

The model of Refs. \cite{BD97, Vol95} is based on the following Hamiltonian\begin{equation}
\label{H0F}
H=H_{0}+F(x,y,z;t),
\end{equation}
 where \( H_{0} \) is the anharmonic oscillator describing the intrinsic
motion of the projectile,

\begin{equation}
\label{equacao}
H_{0}=\frac{1}{2D}\, \, (p_{x}^{2}+p_{y}^{2}+p_{z}^{2})+\frac{C}{2}\, (x^{2}+y^{2}+z^{2})+\frac{B}{4}(x^{2}+y^{2}+z^{2})^{2},
\end{equation}
 where \( D \) is the mass parameter, \( C \) is the oscillator
strength and \( B \) is the strength of the anharmonicity. Here,
we take the mass parameter to be the reduced mass for the motion of
the protons against the neutrons,\[
D=\frac{NZ}{A}m_{0},\]
 where \( m_{0} \) is the average nucleon mass. The beam is assumed
to be parallel to the \( x- \)axis and the coupling interaction \( F \)
is derived from the Lienard-Wiechert potential \cite{WA79} in the
projectile frame

\begin{equation}
\phi (x,y,z,t)=\frac{Z_{T}e\gamma }{[\gamma ^{2}(x-vt)^{2}+(y-b)^{2}+z^{2}]^{1/2}},
\end{equation}
 were \( Z_{T}e \) is the charge of the target, \( b \) is the impact
parameter, and \( \gamma  \) is the Lorentz factor, \( \gamma =1/\sqrt{1-(v/c)^{2}} \).

To be specific, we study the \( ^{208}Pb+^{208}Pb \) collision at
\( 640\,  \)A\( \cdot  \)MeV. We first solve the Schr\"{o}dinger
equation for the intrinsic motion, described by \( H_{0} \). For
this purpose it is convenient to recast the intrinsic Hamiltonian
into the following equivalent form\begin{equation}
\label{int1}
H_{0}=\hbar \omega \, \left[ \frac{1}{2}\left( \pi ^{2}+\rho ^{2}\right) +\beta \, \rho ^{4}\right] \, .
\end{equation}
 In the above, the commonly used variable transformations \begin{equation}
\label{changes}
\rho _{i}=\sqrt{\frac{D\omega }{\hbar }}r_{i};\quad \quad \pi _{i}=\frac{p_{i}}{\sqrt{D\hbar \omega }}\, 
\end{equation}
 have been made, where \( r_{i} \) and \( p_{i} \) stand for the
components of the position and momentum operators respectively. The
oscillator frequency is given by\begin{equation}
\label{hw}
\hbar \omega =\hbar \sqrt{\frac{C}{D}}\, ,
\end{equation}
 and the dimensionless strength \( \beta  \) is related to \( B \)
as\begin{equation}
\label{beta}
B=\left[ \frac{4\, \left( \hbar \omega \right) ^{3}\, D^{2}}{\hbar ^{4}}\right] \, \beta \, .
\end{equation}

In Fig. 1, we show the ratios \( E_{DGDR}^{l=0}\, /\, (2\, E_{GDR}) \)
and \( E_{DGDR}^{l=2}\, /\, (2\, E_{GDR}) \) as a function of \( B, \)
in the same range as chosen in Ref. \cite{BD97}. In this range, the
anharmonicity can be treated using first order perturbation theory
to great accuracy (\( \sim  \) 2\%). The GDR and DGDR energies, to
first order in \( \beta  \), are given by\begin{eqnarray}
E_{GDR}(\beta ) & = & \hbar \omega \, \left( 1+5\, \beta \right) ,\label{EGDR} \\
E_{DGDR}^{l=0}(\beta ) & = & 2\, \hbar \omega \, \left( 1+7.5\, \beta \right) ,\label{DGDR0} \\
E_{DGDR}^{l=2}(\beta ) & = & 2\, \hbar \omega \, \left( 1+6\, \beta \right) \, .\label{DGDR2} 
\end{eqnarray}
 Fig. 1 is equivalent to that shown in Ref. \cite{BD97} and our results
are essentially identical to theirs.

The reduced transition matrix elements can also be easily calculated
to first order in the parameter \( \beta  \). We find\begin{eqnarray*}
\left\langle GDR\left\Vert E1\right\Vert GS\right\rangle  & = & e\, \left( \frac{S_{1}}{\hbar \omega }\right) ^{1/2}\, \left( 1-2.5\, \beta \right) ,\\
\left\langle DGDR,l =0\left\Vert E1\right\Vert GDR\right\rangle  & = & e\, \left( \frac{S_{1}}{\hbar \omega }\right) ^{1/2}\sqrt{\frac{2}{3}}\, (1-5\, \beta ),\\
\left\langle DGDR,l =2\left\Vert E1\right\Vert GDR\right\rangle  & = & e\, \left( \frac{S_{1}}{\hbar \omega }\right) ^{1/2}\sqrt{\frac{10}{3}}\, \left( 1-3.5\beta \right) ,
\end{eqnarray*}
 where \( e \) is the absolute value of the electron charge and \( S_{1} \)
is given by the energy-weighted sum rule,\[
S_{1}=\frac{9}{4\pi }\frac{\hbar ^{2}}{2m_{0}}\frac{NZ}{A}.\]
 The energy-weighted sum rule for transitions from the ground state
and from the GDR are satisfied to first order in the parameter \( \beta  \),
using the above energies and reduced matrix elements.

In order to maintain \( E_{GDR}(\beta ) \) at the experimental value,
namely \( E_{GDR}(\beta )=E_{GDR}^{\exp } \) (13.4 MeV, in the present
case), the oscillator frequency must be renormalized as \( \beta  \)
is changed. The resulting renormalized frequency, from Eq.(\ref{EGDR}),
is\begin{equation}
\label{hwbeta}
\hbar \omega (\beta )=\frac{E_{GDR}^{\exp }}{\left( 1+5\, \beta \right) }\, .
\end{equation}
 Note that in the \( B \)-range of Fig. 1, the dimensionless parameter
varies in the range \( -0.014<\beta <0.014 \) which yields \( 1.08\, E_{GDR}^{\exp }>\hbar \omega (\beta )>0.93 \)
\( E_{GDR}^{\exp } \). Whereas our oscillator frequency is a function
of the anharmonicity parameter, in Ref. \cite{BD97} it is kept constant
at the harmonic value, \( \hbar \omega (\beta =0)=E_{GDR}^{\exp } \).
This difference does not affect the ratio \( E^{l}_{DGDR}/(2\, E_{GDR}) \)
shown in Fig. 1, since the oscillator frequency cancels out in this
case (see eqs. (\ref{EGDR}) to (\ref{DGDR2})). When the renormalized
frequency is used in both the GDR and DGDR energies and matrix elements,
the sum rules for transitions from the ground state and from the GDR
are still satisfied. However, use of the renormalized frequency substantially
changes the excitation probability of the DGDR, as will be shown below.

The calculation of electromagnetic excitation probabilities and cross
sections is performed with the code RELEX\cite{Be99}, based on the
Winther and Alder theory\cite{WA79}. With this code, we perform a
full coupled-channels calculation of the electromagnetic excitation
of the GDR and DGDR. Similar results (about 10\% larger) would be
obtained when perturbation theory is used for the collision dynamics\cite{Be96}.
In Fig. 2, we show the enhancement of the DGDR excitation probability
relative to its harmonic value as a function of \( B \) for the impact
parameter \( b=30 \) fm. We find that for \( B\sim -100 \) MeV/fm\( ^{4} \)
(which in this case corresponds to \( \beta \sim -0.7 \) \( \times 10^{-2} \))
the overall enhancement is 6\%. For purposes of comparison, we have
also performed calculations using a constant frequency (\( \hbar \omega =13.4 \)
MeV in this case). We then obtain an enhancement of 35\%, as shown
by the dashed in line in Fig. 2, in agreement with Ref. \cite{BD97}
(see their Fig. 1).

In Fig. 3a, we show the enhancement in the impact-parameter integrated
DGDR cross section (solid line) vs. \( B \), for the same system.
In the cross section calculations, impact parameters up to 200 fm
are taken into account and a lower cut-off at 15 fm is used to eliminate
nuclear effects. The full line in Fig. 3a represents the result of
the present work, in which an enhancement of only \( 4\% \) is obtained
for \( B=-100 \) MeV/fm\( ^{4} \). The dashed line, obtained using
a fixed value of the oscillator frequency, yields an enhancement of
the DGDR cross section of 22\% for the same value of \( B \). The
\( GDR \) cross section ratio \( \sigma _{GDR}(B)/\sigma _{GDR}(B=0) \)
obtained with fixed GDR energy, shown as a solid line in Fig. 3b,
is close to one over the entire range of \( B \) values but is slightly
less than one for large, negative anharmonicities, (about \( -0.5\% \)
at \( B=-100 \) MeV/fm\( ^{4} \)). This small deviation is due to
the increase in the population of the DGDR at these values of \( B \)
and the correponding depopulation of the GDR. The GDR cross section
ratio obtained with fixed oscillator frequency is shown as a dashed
line in Fig. 3b. In this case, we find the GDR cross section to be
enhanced by about 10\% at \( B=-100 \) MeV/fm\( ^{4} \). The enhancement
of 10\% in the GDR cross section of Fig. 3b is clearly responsible
for the large enhancement of \( 22\% \) in the DGDR cross section
of Fig. 3a at \( B=-100 \) MeV/fm\( ^{4} \) .

The above conclusions do not change noticeably when the calculations
are extended to other systems, such as \( ^{136} \)Xe + \( ^{208} \)Pb
at \( 700\,  \)A\( \cdot  \)MeV. The microscopic study of Ref. \cite{PB00}
established that the anharmonicity parameter scales as \( A^{-1} \)
with the mass number. Thus, if \( B=-100 \) MeV/fm\( ^{4} \) represents
a reasonable value for \( ^{208} \)Pb, then for \( ^{136} \)Xe a
corresponding value would be \( B=-150 \) MeV/fm\( ^{4}. \) In Fig.
4, we display the results of calculations for this system as a function
of the anharmonicity parameter \( B \) in Fig. 4. The solid line
in the figure again shows the results of calculations in which the
oscillator frequency is varied to maintain the GDR energy constant,
while the dashed line represents the results of calculations in which
the oscillator frequency is maintained fixed. Similar to the previous
case, we find the enhancement of the DGDR cross section to be greatly
reduced when the GDR resonance energy is maintained at a fixed value.
As can be seen in Fig. 4, at \( B=-150 \) MeV/fm\( ^{4} \), the
DGDR cross section is enhanced by 62\% when the oscillator frequency
is maintained constant, but is enhanced by less than 10\% when the
GDR energy is maintained at its physical value.

Before ending we comment briefly on the connection between the 
Bortignon-Dasso model used in this paper and microscopic models
\cite{PB00,Cat89,Ber97,Ber99} that aim to assess the importance
of the anharmonic effects both on the spectrum and on the
transition operator. Ref. \onlinecite{Ber97} finds, within
the Lipkin model, small effects on the spectrum (which scale roughly as 1/A).
Hamamoto finds, within nuclear field theory, that the nonlinear effects
in the 1-phonon to 2-phonon transition operator are also quite small
and scale as 1/A\cite{Ham99}. As mentioned above, Ref. \onlinecite{PB00},
through detailed microscopic calculations, finds that the anharmonic
effects are indeed small and scale as 1/A. The values of the parameter
$B$ in both the Bortignon-Dasso and present calculations are taken to be
small enough to be in line with the microscopic findings but also with the 
experimentally observed DGDR excitation energies (although the enhancement of
the DGDR cross section could be increased thorough an artificially large $B$,
there is no choice for this
parameter that whould simultaneously explain the observed cross section 
enhancement and the only very small deviations of the DGDR excitation energy
from the harmonic limit). 

Another interesting point to mention is that the GDR has a width, which
is considered neither by Bortignon and Dasso nor in the present calculation. 
The effect of the
width of the GDR on the excitation of the DGDR has been recently studied
within a harmonic picture\cite{Bert99}. The overall effect of the width,
at the energies considered here is, to produce a slight increase in the
DGDR cross section, although not enough to explain all the available data. 
It  would certainly be of interest to extend the present calculation within the
anharmonic model by coupling the oscillator to other degrees of freedom
(which would generate the damping width).

In conclusion, we have investigated the effect of anharmonicities
in the excitation of the DGDR in relativistic heavy ion collisions,
with the same macroscopic model used by Bortignon and Dasso\cite{BD97}.
We point out that variations of the anharmonicity strength must be
accompanied by a renormalization of the oscillator frequency, in order
to maintain the GDR energy at a value consistent with the experimental
one. We have found that this condition strongly reduces the enhancement
in the DGDR excitation probabilities and corresponding cross sections,
so that they remain much below the experimental results.

\bigskip

This work was supported in part by DAAD/CAPES cooperative agreement
no. 415-bra-probral/bu, CNPq and the MCT/FINEP/CNPq(PRONEX) under
contract no. 41.96.0886.00. D.T.P and L.F.C. acknowledge partial support
from the Funda\c{c}\~{a}o Universit\'{a}ria Jos\'{e} Bonif\'{a}cio,
and M.S.H. and B.V.C. acknowledge support from the FAPESP. 
\textbf{Figure Captions}

\begin{itemize}
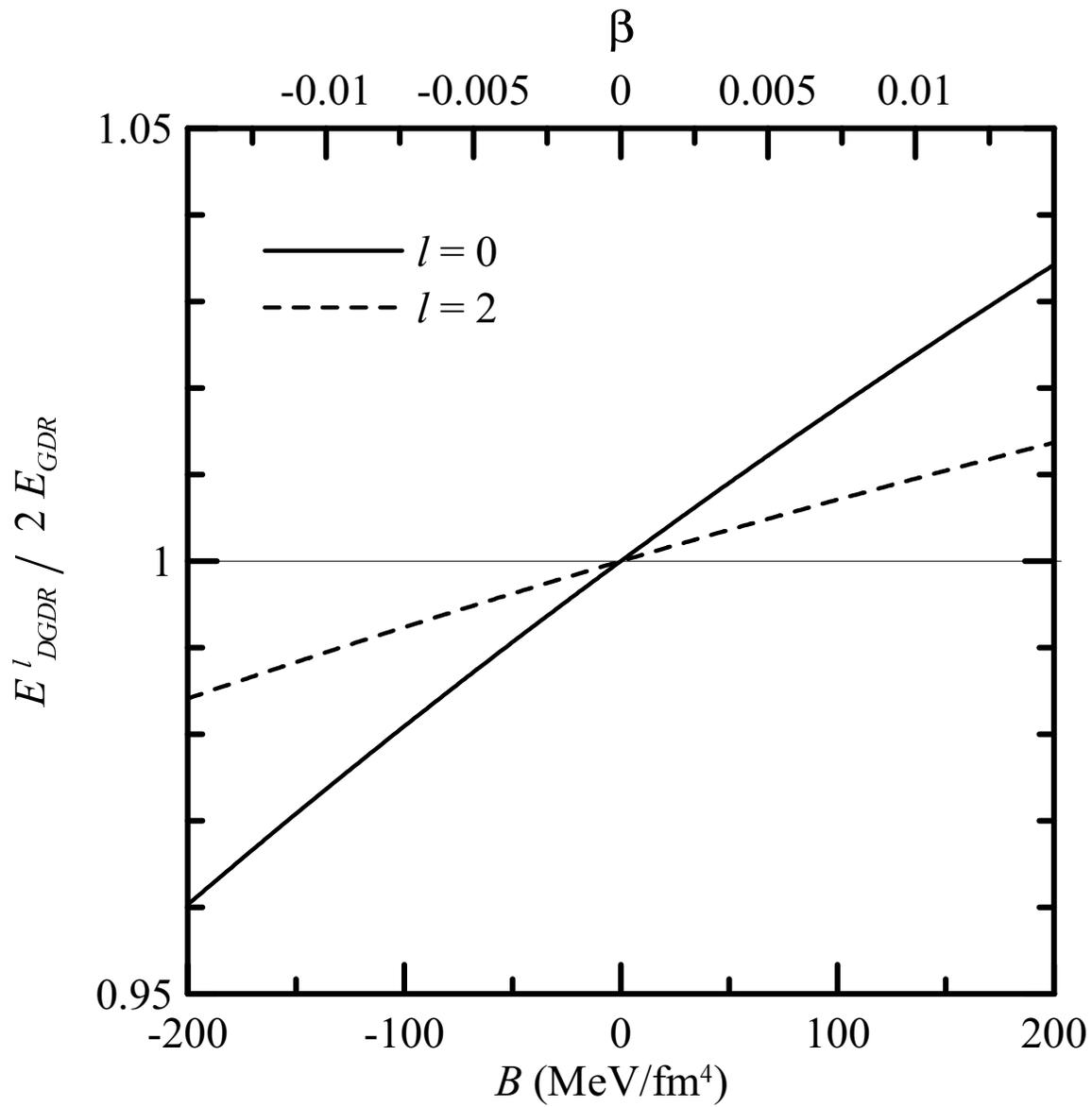
\item Figure 1: The ratio \( E_{DGDR}^{l}\, /\, (2\, E_{GDR}) \) vs the
anharmonicity parameter \( B, \) for \( ^{208}Pb \). The solid line
is for \( l=2 \) and the dashed line for \( l=0. \) The reduced
mass for the oscillation of protons against neutrons is used for the
mass parameter \( D \). 
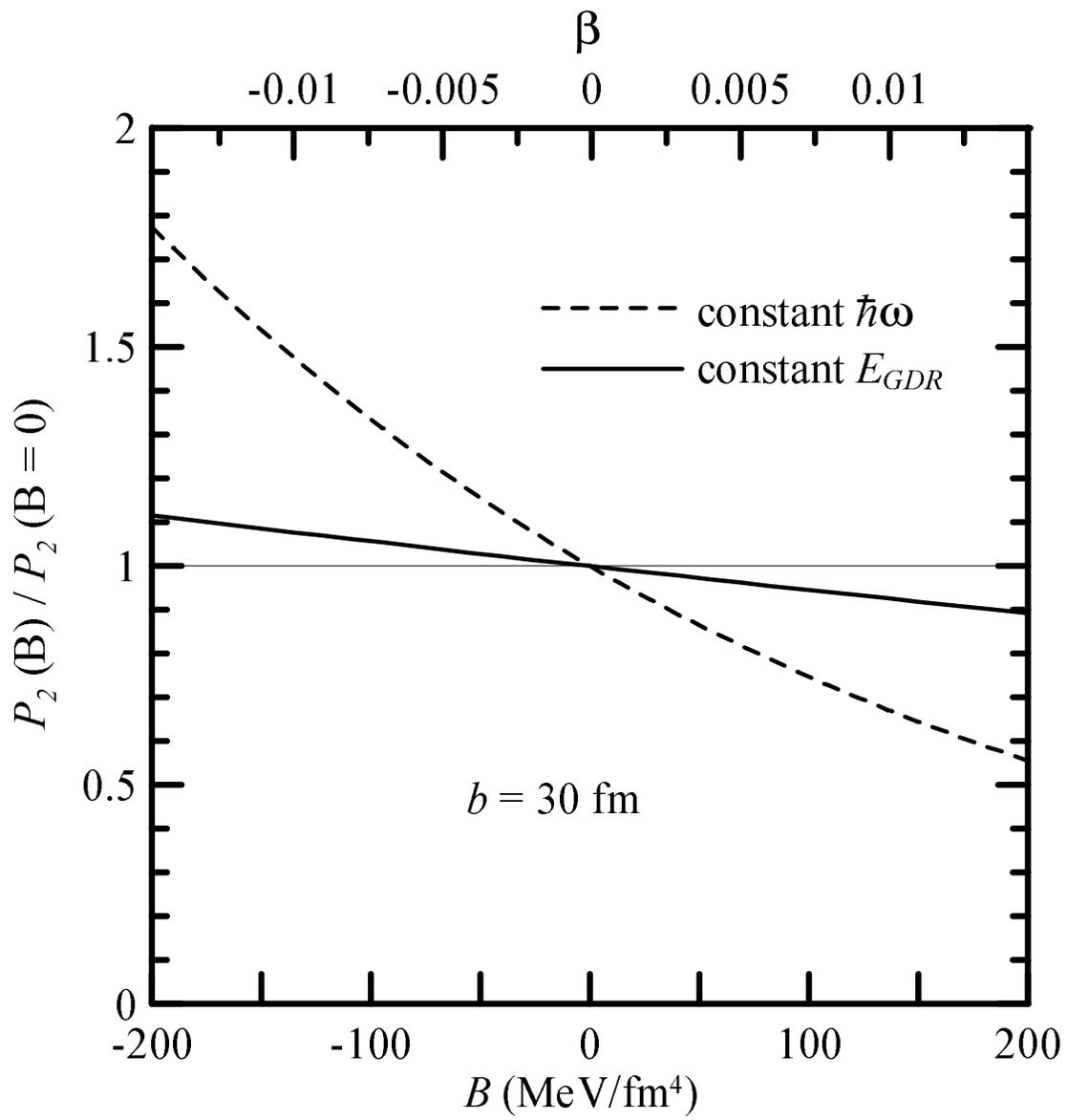
\item Figure 2: The enhancement in the excitation of the DGDR in the collision
of \( ^{208}Pb+^{208}Pb \) at \( 640\,  \)A\( \cdot  \)MeV for
the impact parameter \( b=30 \) fm. The solid line represents the
results of the present calculation while the dashed line corresponds
to a constant oscillator frequency. 
\item Figure 3: Enhancement factor of the (a) DGDR and (b) GDR cross sections
in the collision of \( ^{208}Pb+^{208}Pb \) at \( 640\,  \)A\( \cdot  \)MeV.
The dashed lines correspond to the results obtained with fixed oscillator
frequency, while the full lines correspond to a fixed \( E_{GDR} \). 
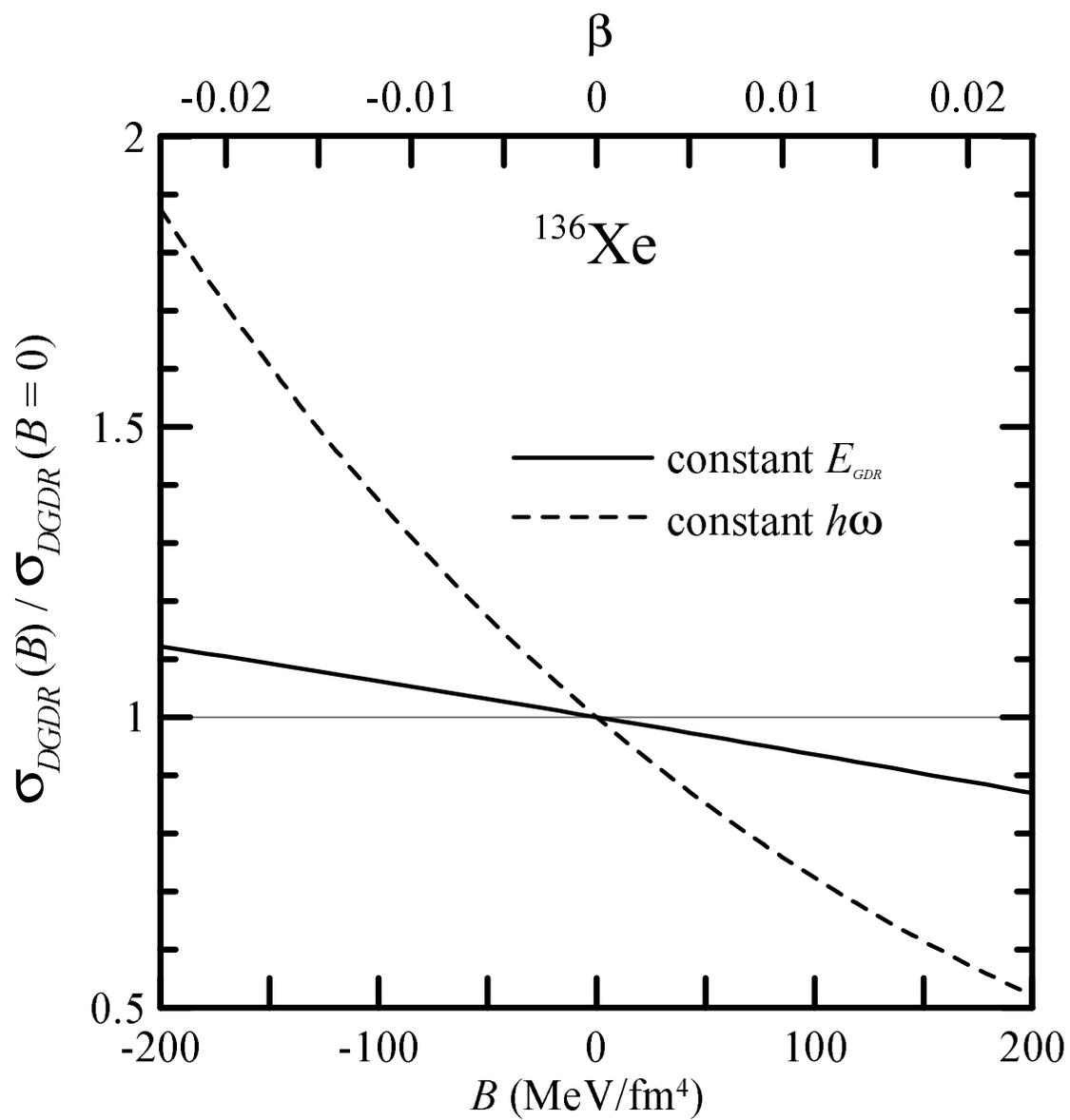
\item Figure 4: Enhancement factor of the DGDR cross section in the collision
of \( ^{136}Xe+^{208}Pb \) at \( 700\,  \)A\( \cdot  \)MeV. The
dashed line corresponds to the results obtained with fixed oscillator
frequency, while the full line corresponds to a fixed \( E_{GDR} \). \end{itemize}

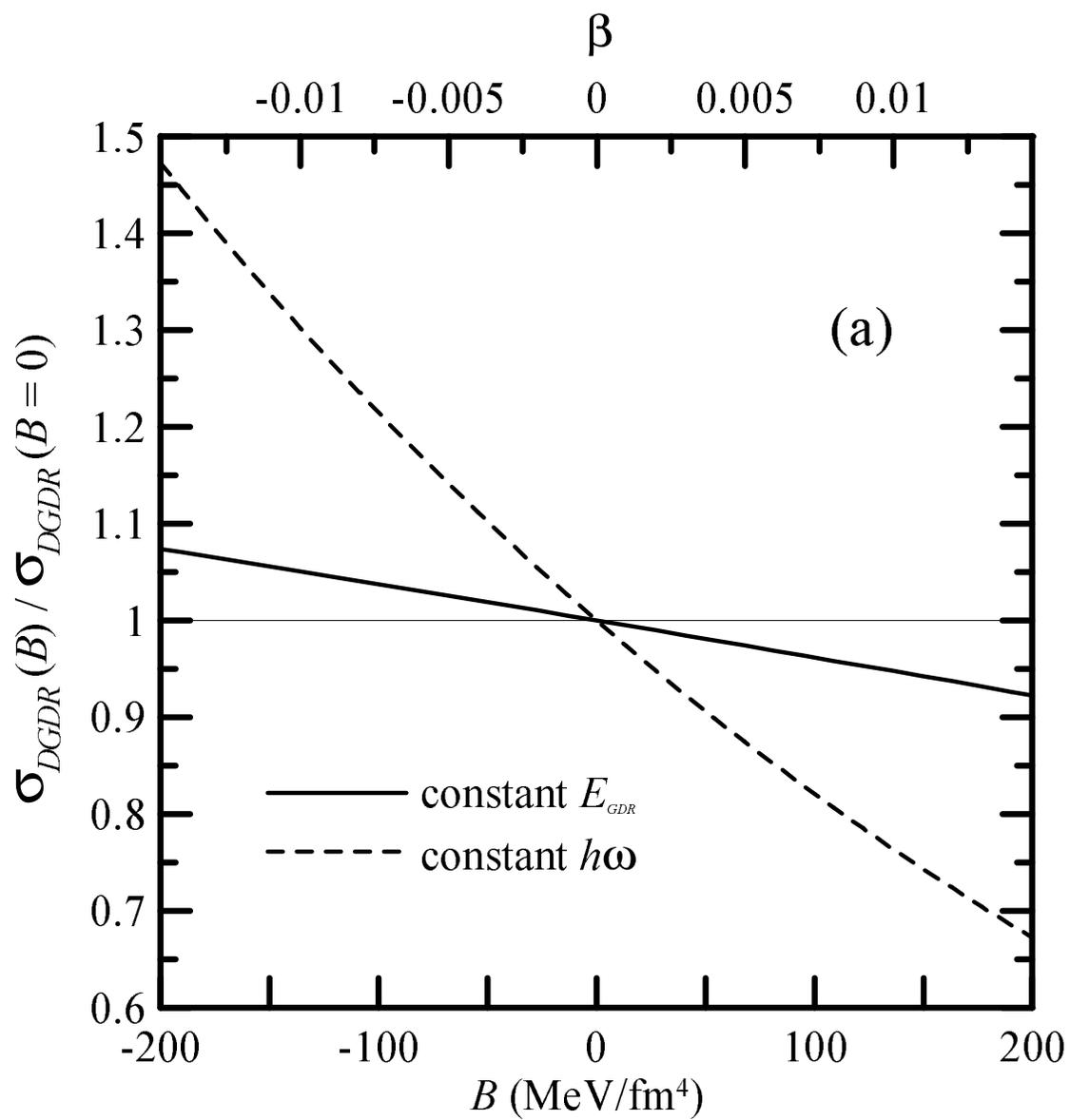
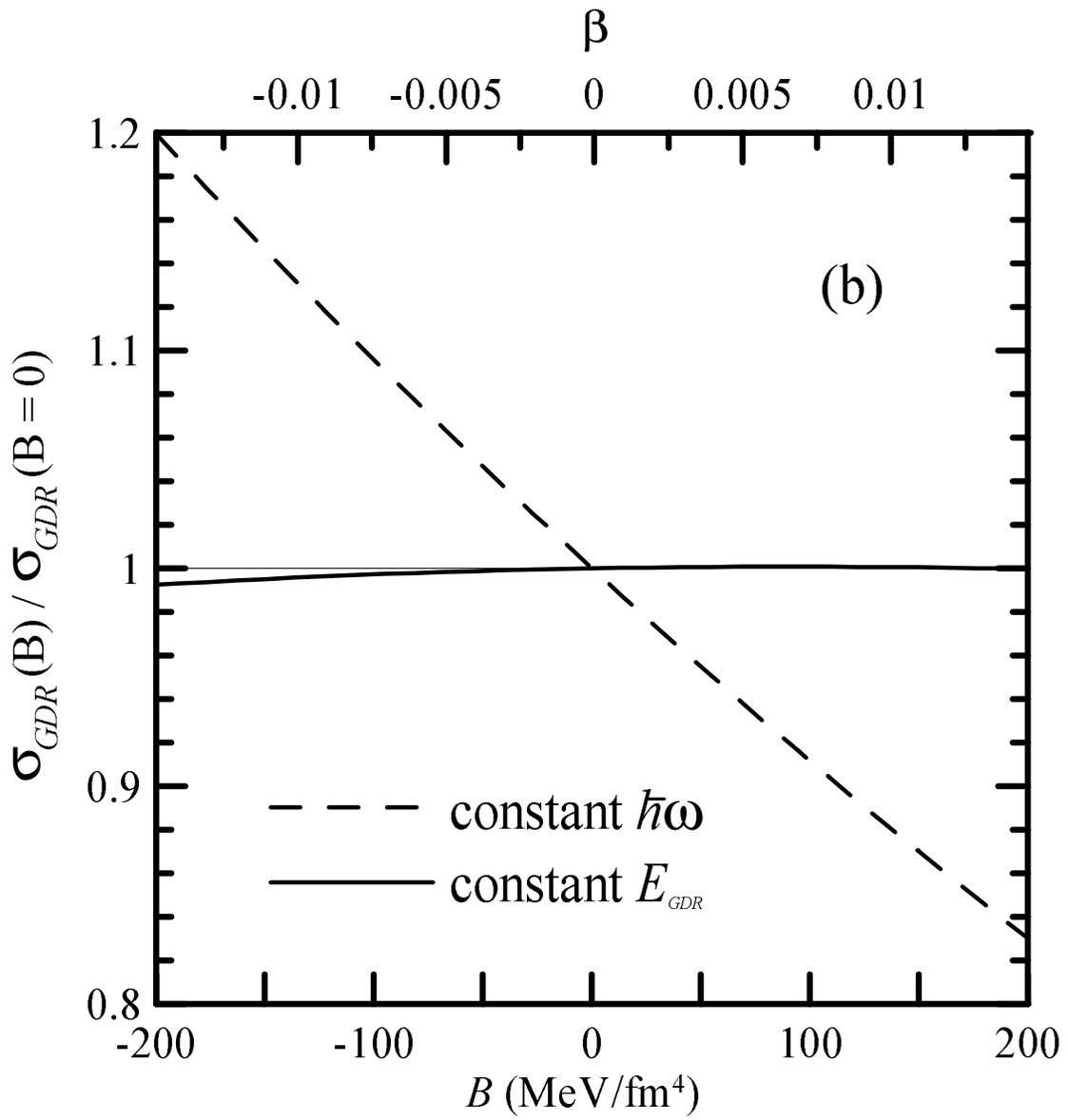
\end{document}